\documentclass[aps, prd,twocolumn,aps,amsmath,nofootinbib,superscriptaddress, preprintnumbers]
{revtex4}

\usepackage{graphicx}
\usepackage{hyperref}
\usepackage{bm}

\newcommand{\nn}{\nonumber} 
\newcommand{\bea}{\begin{eqnarray}}
\newcommand{\eea}{\end{eqnarray}}
\newcommand{\beq}{\begin{equation}}
\newcommand{\eeq}{\end{equation}}
\newcommand{\comment}[1]{}

\newcommand{\mR}{{\mathcal R}}

\begin{document}


\preprint{MIT-CTP-4098} 

\title{The distribution of $\Omega_k$ from the scale-factor cutoff measure}

\author{Andrea De Simone}
\affiliation{Center for Theoretical Physics, Laboratory for Nuclear 
Science, and Department of Physics, \\
Massachusetts Institute of Technology, Cambridge, MA 02139}

\author{Michael P.~Salem}
\affiliation{Institute of Cosmology, Department of Physics and Astronomy,\\
Tufts University, Medford, MA 02155}


\begin{abstract}
Our universe may be contained in one among a diverging number 
of bubbles that nucleate within an eternally inflating 
multiverse.  A promising measure to regulate the diverging 
spacetime volume of such a multiverse is the scale-factor 
cutoff, one feature of which is bubbles are not rewarded for 
having a longer duration of slow-roll inflation.  Thus, 
depending on the landscape distribution of the number of 
e-folds of inflation among bubbles like ours, we might hope 
to measure spacetime curvature.  We study a recently proposed
cartoon model of inflation in the landscape and find a reasonable 
chance (about ten percent) that the curvature in our universe is 
well above the value expected from cosmic variance.  Anthropic 
selection does not strongly select for curvature as small as is 
observed (relative somewhat larger values), meaning the 
observational bound on curvature can be used to rule out landscape 
models that typically give too little inflation.
\end{abstract}

\pacs{98.80.Cq}

\maketitle

\section{Introduction}
\label{sec:introduction}

Inflation is generically eternal, with the physical volume of 
false-vacuum inflating regions increasing exponentially with time 
and ``pocket universes'' like ours constantly nucleating out of 
the false vacuum.  Each of these pockets contains an infinite, 
open Friedmann--Robertson--Walker (FRW) universe and, when the 
fundamental theory admits a landscape of meta-stable vacua, each 
may contain different physical parameters, or even different 
fundamental physics, than those observed within our universe.  
In order to make meaningful predictions on what physics we should 
expect to observe within our pocket it is necessary to adopt a 
prescription  to regulate the diverging spacetime volume  of the 
multiverse (for recent reviews, see for example 
Refs.~\cite{Winitzki06,Guth07,Linde07,AV06,AGJ}).  

This issue, known as the measure problem, has been  addressed in 
several different ways so far~\cite{LM,LLM,GBLL,AV94,AV95,GTV,
pockets,GSPVW,ELM,diamond,censor,Sinks,LindeGI,Vanchurin07,
Winitzki08,GV08b,LindeGI2,Bousso09}.  Different approaches in 
general make different observational predictions, and some of these 
apparently conflict with observation.  For example, approaches 
that define probabilities with respect to a global proper-time 
foliation~\cite{LM,LLM,GBLL,Sinks,AV94} suffer a ``youngness 
paradox,'' predicting that we should have evolved at a very 
early cosmic time, when the conditions for life were very
hostile~\cite{youngness,BFY}.  Volume-weighting measures, like 
the so-called ``gauge invariant'' or ``stationary'' 
measures~\cite{AV95,LindeGI,LindeGI2} and the pocket-based 
measures~\cite{GTV,pockets,GSPVW,ELM}, have a ``runaway 
inflation'' problem.  These measures predict that we should 
observe severely large or small values of the primordial 
density contrast~\cite{FHW,QGV} and the gravitational 
constant~\cite{GS}, while these parameters appear to sit 
comfortably near the middle of their respective anthropic 
ranges~\cite{anthQ,GS}.  (Some suggestions to possibly get 
around this issue have been described in Refs.~\cite{QGV,HWY}.)  

The causal patch measure~\cite{diamond,censor} and the 
scale-factor cutoff measure~\cite{DSGSV1,BFYSFC} survive 
these hazards.  Furthermore, under reasonable assumptions about 
the landscape~\cite{diamond,DSGSV2}, these measures do not 
suffer a ``Boltzmann brain invasion''~\cite{Rees1,Albrecht,
DKS02,Page1,Page2,Page06,BF06}, where observers created as rare 
quantum fluctuations outnumber ``normal'' observers who evolve 
from out-of-equilibrium processes in the wake of the big bang.  
There is also encouraging evidence that these measures coincide 
with probability measures stemming from independent theoretical 
considerations~\cite{GV08b,Bousso09}.  Thus we consider these
measures to be promising proposals to regulate the diverging
spacetime volume of the multiverse.         

An interesting feature of these measures  is that they do not 
reward regions of the multiverse for having longer periods of 
slow-roll inflation.  Thus, one might hope that in our bubble 
slow-roll inflation did not last long enough to wash away all of 
the relics of the bubble nucleation event.  One such relic is 
the large geometric curvature of the bubble at the time of its 
nucleation.  (Note that the large-curvature initial bubble 
is still homogeneous and isotropic, due to the symmetries of 
the eternally-inflating vacuum in which it 
nucleates~\cite{CdL,LW}.)  In this paper, we study the probability 
distribution  of the curvature parameter 
$\Omega_k$,
\bea
\Omega_k = 1 - \frac{8\pi G}{3H_0^2} \rho_{\rm total} \,,
\eea
where $H_0$ is the Hubble parameter today and $\rho_{\rm total}$ is
the total energy density of our universe.

For simplicity, we only focus on the scale-factor cutoff measure. 
The joint probability distribution for $\Omega_k$ and the 
cosmological constant, using the causal entropic 
principle~\cite{cep}, has been already investigated in 
Ref.~\cite{bozek}.  The predictions of the two approaches turn 
out to be very similar.  Because the causal patch measure has 
overlapping features with the casual entropic principle and the
scale-factor cutoff measure (see e.g.~Ref.~\cite{BFYSFC}), we expect
it to also give similar results.

We first study the effect of anthropic selection in favor of 
small $\Omega_k$, which derives from the tendency of large 
curvature to inhibit structure formation.  Anthropic 
distributions for the curvature parameter have previously 
been estimated by Vilenkin and Winitzki~\cite{VW} and by 
Garriga, Tanaka, and Vilenkin~\cite{GTV}; however that work 
did not include a non-zero cosmological constant.  The 
cosmological constant was included in a more recent 
calculation by Freivogel, Kleban, Rodriguez Martinez, and 
Susskind ~\cite{FKMS}; however that work did not take 
into account the Gaussian distribution of primordial density 
perturbations, which allows for structure formation even 
when the curvature is large enough to prevent the collapse of 
a typical-amplitude density fluctuation.  We provide a 
complete treatment of the problem, using updated cosmological 
data.  Although anthropic selection strongly suppresses the 
probability to measure $\Omega_k>0.6$ or so, by itself it does 
not strongly select for values of $\Omega_k$ as small as the 
present observational bound.

The curvature parameter $\Omega_k$ depends exponentially on 
the number of e-folds of slow-roll inflation $N_e$.  The authors of 
Ref.~\cite{FKMS}  proposed a simple toy model of 
inflation in the landscape, for which they find $N_e$ to 
follow the distribution $dP_0(N_e)\propto N_e^{-4}dN_e$.  We 
adopt this model and use the scale-factor cutoff measure to 
predict the distribution of $\Omega_k$ among bubbles like 
ours in the multiverse.  The result is essentially what one 
might guess by ignoring volume weighting, anthropic selection, 
and the effects of the measure.  The predicted distribution 
of $\Omega_k$ prefers values below that expected from cosmic 
variance~\cite{WZ,VTS}, but it gives reasonable  hope for 
$\Omega_k$ to be significantly larger.  Specifically, there is 
about a 6\% chance to observe $\Omega_k\geq 10^{-3}$ and about 
an 11\% chance to observe $\Omega_k\geq 10^{-4}$, the latter
corresponding roughly to accuracy to which $\Omega_k$ can in
principle be determined~\cite{VTS}.  These predictions rely on 
some simple assumptions about inflation, including a reheating 
temperature of $T_*\approx 10^{15}$ GeV.  (All else being equal, 
lowering the reheating temperature increases the likelihoods 
for these observations.)

To make the above predictions as precise as possible, we have
assumed that $\Omega_k$ is measured at the present cosmic 
time, and input the observational constraint 
$\Omega_k\leq 0.013$~\cite{WMAP5} (for simplicity we treat 
this 95\% confidence level as a hard bound).  Yet, related
to the question of what we (observers living at the present 
cosmic time) expect to measure, there is the question of 
what typical observers (i.e. those living at arbitrary times) 
in bubbles like ours measure.  To address this question it is
convenient to work with a time-invariant measure of 
curvature; for this we choose  
\bea
k=\left(\frac{\Omega_k^3}{\Omega_\Lambda\Omega_m^2}\right)^{\!1/3} ,
\eea
which in effect expresses the inverse curvature radius squared,
$r_{\rm curv}^{-2}=H^2\Omega_k$, in units related to the 
late-time matter density and cosmological constant (here 
$\Omega_\Lambda$ is the density parameter of the cosmological 
constant,  $\Omega_m$ is that of non-relativistic matter). 
As before we restrict attention to bubbles just like ours, including
the value of the cosmological constant, and times after 
non-relativistic matter domination, when presumably all of the
observers arise.  One can then ask how typical is our 
measurement, $k\leq 0.035$. Using the scale-factor cutoff, we 
find that observers typically observe $k$ to satisfy this bound.   

Because anthropic selection is rather weak in the 
vicinity of the bound $k\leq 0.035$, we can rule out 
certain distributions of $N_e$, because they predict that we 
should measure $k$ to be much larger than we do.  The 
assumptions referred to above relate $k=0.035$ to $N_e=63.7$ 
e-folds of inflation.  Although anthropic selection is weak 
for $N_e$ near to and greater than this number, it becomes 
strong at $N_e\approx 61$.  Thus, a landscape distribution 
of $N_e$ can be ruled out if its weight over the interval 
$63.7\leq N_e$ is much less than its weight over the 
interval $61\lesssim N_e < 63.7$.  Different assumptions about 
inflation (for example higher or lower reheating temperature) 
merely shift the numbers in these inequalities.  

The remainder of this paper is organized as follows.  In 
Section~\ref{sec:background}, we review some background 
material that is relevant to our main calculation, including 
a brief description of the scale-factor cutoff measure 
(Section~\ref{sec:sfcutoff}), a description of how one can 
model bubble geometry before and after slow-roll inflation
(Section~\ref{sec:geometry}), and some background on 
structure formation in an open FRW universe 
(Section~\ref{sec:structure}).  The distribution of 
$\Omega_k$ is calculated in Section~\ref{sec:omega}.  In 
Section~\ref{sec:anth} we discuss anthropic considerations 
and describe how our results can be used to rule out 
hypothetical models of inflation in the landscape.  The 
analysis of Sections~\ref{sec:omega} and~\ref{sec:anth} is 
discussed in the context of an alternative form of the 
scale-factor cutoff measure in  Appendix~\ref{sec:altSFC}, 
where it is shown the predictions are qualitatively unchanged.  
We draw our conclusions in Section~\ref{sec:conclusions}.

\section{Background}
\label{sec:background}

\subsection{The Scale Factor Cutoff Measure}
\label{sec:sfcutoff}

Perhaps the simplest way to regulate the infinities of eternal
inflation is to impose a cutoff on a hypersurface of constant
global time~\cite{LM,LLM,GBLL,AV94,AV95}.  One starts with a
patch of a spacelike hypersurface $\Sigma$ somewhere in an
inflating region of spacetime, and follows its evolution along
the congruence of geodesics orthogonal to $\Sigma$.  The
scale-factor time is defined as
\beq
t=\ln a \,,
\label{tdef}
\eeq
where $a$ is the expansion factor along the geodesics.  The 
scale-factor time is related to the proper time $\tau$ by
\beq
dt = H\,d\tau \,,
\label{ttau}
\eeq
where $H$ is the Hubble expansion rate of the congruence.  The
spacetime region swept out by the congruence will typically expand 
to unlimited size, generating an infinite number of pockets.  
(If the patch does not grow without limit, one chooses another 
initial patch $\Sigma$ and starts again.)  The resulting
four-volume is infinite, but we cut it off at some fixed
scale-factor time $t=t_c$.  To find the relative probabilities of
different events, one counts the numbers of such events in the
finite spacetime volume between $\Sigma$ and the
$t=t_c$ hypersurface,
and then takes the limit $t_c\to\infty$. 

The term ``scale factor'' is often used in the context of homogeneous
and isotropic geometries; yet on very large and on very small scales 
the multiverse may be very inhomogeneous.  A simple way to deal with 
this is to take the factor $H$ in Eq.~(\ref{ttau}) to be the local
divergence of the four-velocity vector field along the congruence of
geodesics orthogonal to $\Sigma$, 
\beq
H(x) \equiv (1/3) \,u^\mu_{\phantom{\mu};\,\mu} \,.
\label{localsft}
\eeq
When more than one geodesic passes through a point, the scale-factor
time at that point may be taken to be the smallest value among the set 
of geodesics.  In collapsing regions $H(x)$ is negative, in which case 
the corresponding geodesics are continued unless or until they hit a 
singularity.

This ``local'' definition of scale-factor time has a simple geometric 
meaning.  The congruence of geodesics can be thought of as representing 
a ``dust'' of test particles scattered uniformly on the initial 
hypersurface $\Sigma$.  As one moves along the geodesics, the density 
of the dust in the orthogonal plane decreases.  The expansion factor 
$a$ in Eq.~(\ref{tdef}) can then defined as $a\propto\rho^{-1/3}$, 
where $\rho$ is the density of the dust, and the cutoff is triggered 
when $\rho$ drops below some specified level.

Although the local scale-factor time closely follows the FRW scale 
factor in expanding spacetimes --- such as inflating regions and 
thermalized regions not long after reheating --- it differs dramatically 
from the FRW scale factor as small-scale inhomogeneities develop during 
matter domination in universes like ours.  In particular, the local 
scale-factor time nearly grinds to a halt in regions that have decoupled 
from the Hubble flow.  It is not clear whether we should impose this 
particular cutoff, which would essentially include the entire lifetime 
of any nonlinear structure that forms before the cutoff, or impose a 
cutoff on some nonlocal time variable that more closely tracks the FRW 
scale factor.\footnote{The distinction between these two forms of 
scale-factor time was first pointed out by Bousso, Freivogel, and Yang 
in Ref.~\cite{BFYSFC}.} 

Note however that if we focus on an idealized multiverse composed 
entirely of thin-wall bubbles, in which bubble collisions do not 
significantly deform one of the involved bubbles, then it is possible 
to unambiguously define the FRW Hubble rate $H$ at any point along the 
congruence.  In particular, the initial FRW symmetry of each bubble defines a 
foliation over which the expansion rate (\ref{localsft}) can be 
spatially averaged.  Scale-factor time is continued from one bubble to 
another by taking it to be continuous across bubble walls.  We 
emphasize that for more general landscapes, such a definition is not 
available.  We leave to future work developing a nonlocal modification 
of the scale-factor time (\ref{localsft}) that both approximates our 
intuitive notion of FRW averaging and also extends into more 
complicated geometries.

In general, the local scale-factor cutoff measure defined by 
(\ref{ttau}) and (\ref{localsft}) and the nonlocal scale-factor cutoff 
defined above make different predictions for physical observables.   
In the present case of the curvature parameter, however, the prediction 
of each choice is qualitatively the same.  In the main body of this 
paper we refer to the above nonlocal definition of scale-factor time, 
for which we take the FRW scale factor as a suitable
approximation.  In Appendix~\ref{sec:altSFC} the analysis is briefly 
repeated using the local scale-factor time parametrization.


To facilitate further discussion, we here review some general 
features of eternally inflating spacetimes in light of a
scale-factor time foliation.  Regions of an eternally 
inflating multiverse might involve fields sitting in local 
potential minima, at least some of which correspond to 
positive false-vacuum energies.  Evolution of the multiverse 
is then governed by bubble nucleation through quantum 
tunneling~\cite{Gott,Steinhardt}, either from one local 
minimum to another or from a local minimum to a region of 
of classical slow-roll inflation.\footnote{\label{quantdiff}Regions 
of the multiverse may also (or instead) be described by 
quantum diffusion~\cite{eternal1,Linde86,starobinsky}, i.e. 
eternal inflation may be driven by the potential energy of 
some light scalar field(s), the evolution of which is 
dominated by quantum fluctuations.  Pockets form when the 
scalar field(s) fluctuate into a region of parameter space 
where classical evolution dominates, and slow-roll inflation 
ensues.  In this case, the pocket geometry in general does 
not contain the global symmetry of those formed via bubble 
nucleation.  Nevertheless, the curvature perturbation at the 
onset of slow-roll inflation is still typically order unity, 
and an analysis similar to that in this paper could be 
performed.  On the other hand, models of inflation featuring 
quantum diffusion typically have very flat potentials and a 
large number of e-folds between the onset and end of 
classical slow-roll inflation.  Therefore, in this paper we 
ignore this type of cosmological evolution.}  
In the latter case, the bubble interiors have the geometry of 
open FRW universes.  Bubbles of interest to us here have a 
period of slow-roll inflation followed by reheating.  

In the limit of large scale-factor time, the number of objects 
of any type that have formed prior to time $t$ is (asymptotically) 
proportional to $e^{\gamma t}$, where $\gamma$ is a universal 
constant.  This universal asymptotic scaling behavior is a 
consequence of the physical volume of the multiverse being 
dominated by a subset of vacua, the ``dominant'' vacua, whose 
collective volume grows at a constant rate.  This is the set 
of de Sitter vacua corresponding to the state with the 
smallest-magnitude eigenvalue in the master rate equation for 
the multiverse~\cite{SPV}.  In many landscapes, there is just 
one dominant vacuum, the de Sitter vacuum with the smallest 
decay rate.  For our purposes it is sufficient to note that
\bea
\gamma \approx 3 \,,
\label{gammascale}
\eea
with corrections on the order of the exponentially suppressed 
decay rate of the dominant vacuum (specifically, the 
smallest-magnitude rate equation eigenvalue).  As an example 
of this asymptotic scaling behavior, the number of bubbles of 
our type that nucleate between the time $t$ and $t+dt$ has 
the form
\bea
dn \propto e^{\gamma t}dt \,.
\label{dn}
\eea
This scaling behavior holds even when the multiverse contains 
regions governed by quantum diffusion (c.f. 
Footnote~\ref{quantdiff}); for details see for example 
Ref.~\cite{WV95}.

\subsection{The Geometry of Pocket Nucleation}
\label{sec:geometry}

We here provide some background on the geometry of bubble 
nucleation; this section is largely based on Ref.~\cite{VW} 
(while this paper was in preparation, a similar analysis 
appeared in Ref.~(\cite{BFYSFC})).  To begin, consider a 
bubble of our vacuum that nucleates at scale-factor time 
$t_{\rm nuc}$.  The parent vacuum in which our bubble 
nucleates can be described by flat de Sitter coordinates 
with metric
\bea
ds^2 = -dt^2 + e^{2t}\left( dr^2+r^2d\Omega_2^2\right) \,, 
\eea  
where $t$ is the flat de Sitter time, defined so as to 
coincide with the scale-factor time in the parent vacuum, 
and $d\Omega_2^2=d\theta^2+\sin^2\theta d\phi^2$.  We 
assume the parent vacuum has larger vacuum energy than 
ours.  The nucleation process is then as described in 
Ref.~\cite{CdL}:  the bubble starts as a small three-sphere 
and expands at a rate that rapidly approaches the speed of 
light.  

Inside the bubble, we take interest in the open FRW 
coordinates $(\tau,\xi)$, which are described by the metric
\bea
ds^2 = -d\tau^2 + \tilde{a}^2(\tau)
\left( d\xi^2+\sinh^2\xi\,d\Omega_2^2\right) \,.
\label{openFRW}
\eea
Here $\tilde{a}(\tau)$ is the scale factor within the 
bubble, which should not be confused with that outside the 
bubble.  We define proper time $\tau$ such that $\tau=0$ at 
the bubble wall.  The coordinates $(\tau,\xi)$ are natural 
to an observer inside the bubble --- surfaces of constant 
proper time $\tau$ have constant energy density and a 
constant curvature term $1/\tilde{a}^2$, i.e. the Einstein 
field equation gives
\bea
H^2-\frac{1}{\tilde{a}^2}=\frac{8\pi G}{3}\,\rho_{\rm total} \,.
\label{EFEpocket}
\eea
Note that curves of constant $\xi$, $\theta$, and $\phi$ 
define a congruence of comoving geodesics inside the 
bubble.

\begin{figure}[t!]
\includegraphics[width=0.4\textwidth]{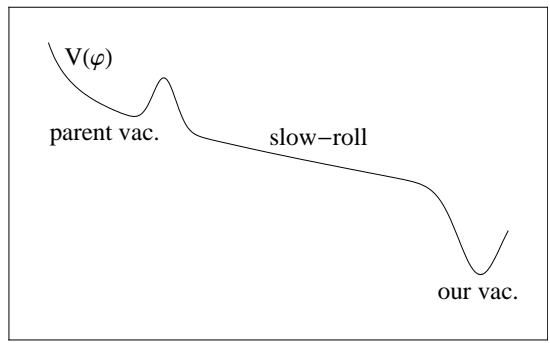} 
\caption{The potential $V(\varphi)$ describing the parent 
vacuum, slow-roll inflation potential, and our vacuum.}
\label{fig:V}
\end{figure}

In order to obtain a simple relationship between the 
global geodesic congruence and that defined inside the 
bubble, we consider a simple model of bubble 
nucleation~\cite{VW}.  Specifically, we model the 
false-vacuum inflation of the parent vacuum, the tunneling 
event, and the subsequent evolution in the bubble (up to 
reheating) using a single scalar field $\varphi$, with 
potential $V(\varphi)$ (as illustrated in Fig.~\ref{fig:V}).  
Furthermore, we assume the tunneling barrier of $V$ is such 
that $V(\varphi)$ is nearly the same before and after 
tunneling, and that gravitational effects of the bubble wall 
are negligible.  Due to the symmetries of the tunneling 
instanton, the field $\varphi$ emerges after tunneling with 
zero `velocity', $d\varphi/d\tau=0$~\cite{CdL}.  Therefore, 
at very early times $\tau$ the geometry inside the bubble 
is approximately de Sitter.       

Because the vacuum energy is nearly the same outside and 
just inside the bubble, and the geometry in both regions is 
de Sitter, constant $r$ geodesics pass unaffected through 
the bubble wall.  Thus, in this de Sitter region the global 
geodesic congruence and that inside the bubble are related 
by the usual relationship between flat and open de Sitter 
coordinates:
\bea
H_i t (\tau,\xi) &=& \ln\big[
\cosh(H_i\tau)+\sinh(H_i\tau)\cosh\xi\,\big] \nn\\
H_i r (\tau,\xi) &=& \frac{\sinh(H_i\tau)\sinh\xi}
{\cosh(H_i\tau)+\sinh(H_i\tau)\cosh\xi} \,,
\label{coordtrans}
\eea
where $H_i$ is the Hubble rate of the parent vacuum.  Note 
that $H_i$ is not the Hubble rate at early times in the 
bubble, even though the energy density $V$ is nearly the 
same in both regions.  This is because of the curvature 
term in Eq.~(\ref{EFEpocket}).  Solving Eq.~(\ref{EFEpocket})
in the limit $V(\varphi)\approx V(\varphi_i) = 3H_i^2/8\pi G$, 
one finds
\bea
\tilde{a}(\tau)=H_i^{-1}\sinh(H_i\tau) \,,
\label{sftau}
\eea   
(the singularity $a\to 0$ as $\tau\to 0$ is only a 
coordinate singularity).  This solution holds as long as 
$V(\varphi)$ does not change significantly, i.e. as long as 
$H_i\tau \ll \sqrt{16\pi G}\, V/V'$, where the prime 
denotes $\varphi$-differentiation~\cite{VW}.  

\begin{figure}[t!]
\includegraphics[width=0.4\textwidth]{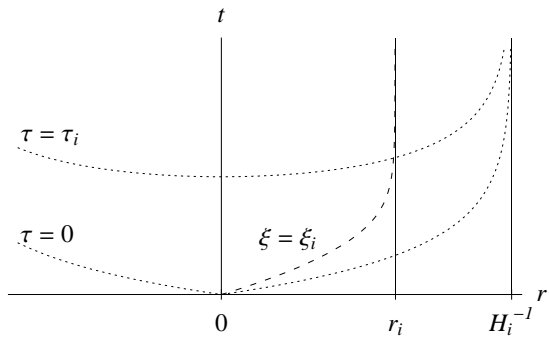} 
\caption{The geometry near the bubble boundary.}
\label{fig:coords}
\end{figure}

After entering the de Sitter region just inside the bubble 
wall, geodesics of constant $r$ (which are comoving in the 
parent vacuum) asymptote to the geodesics of constant $\xi$ 
(which are comoving in the bubble), up to corrections of 
order $e^{-H_i\tau}$.  See Fig.~\ref{fig:coords} for an
illustration.  We assume that we can map these geodesics 
onto each other with reasonable accuracy during the early 
de Sitter expansion, i.e. we assume there exists a time 
$\tau_i$ satisfying 
$1\ll H_i\tau_i \ll \sqrt{16\pi G}\, V/V'$.  The 
scale-factor time at $\tau_i$ is then given by
\bea
t_i=t_{\rm nuc} + H_i\tau_i+2\ln\cosh\left(\xi/2\right) \,,
\label{coords}
\eea
which obtains by taking the limit $H_i\tau_i\gg 1$ of 
Eqs.~(\ref{coordtrans}).

After the proper time $\tau_i$, the bubble expands through 
$N_e$ e-folds of slow-roll inflation, reheats, and then 
undergoes big bang evolution.  We will take interest in a 
reference class of observers who measure $\Omega_k$ at the 
same FRW proper time, $\tau_0$.  We denote the number of 
e-folds of expansion along a constant-$\xi$ geodesic from 
reheating to this time $N_0$.  Then the scale-factor time 
at which these observers measure $\Omega_k$ can be written
\bea
t_0 = t_{\rm nuc} + H_i\tau_i+2\ln\cosh\left(\xi/2\right) 
+ N_e + N_0 \,.
\label{SFO}
\eea
Note that $t_0$ is a function of $\xi$.  Thus, the 
scale-factor cutoff, which requires $t_0\leq t_c$, implies 
a cutoff on the FRW coordinate $\xi$, $\xi\leq\xi_c$, with
\bea
\xi_c = 2\cosh^{-1}\! 
\exp\!\left[\frac{1}{2}\left(t_c-t_{\rm nuc}-H_i\tau_i
- N_e - N_0\right)\right] \! .\,\,\,\,
\label{xic}
\eea
The cutoff $\xi_c$ in turn implies that the constant 
$\tau=\tau_0$ hypersurface on which $\Omega_k$ is measured, 
$\Sigma_0$, is finite.

The number of observers that arise in the bubble before 
the cutoff $t_c$ is proportional to the physical volume of 
$\Sigma_0$.  More precisely, the number of observers is 
proportional to the matter density on $\Sigma_0$ times its 
physical volume.  After inflation the matter density dilutes 
with cosmic expansion, so the number of observers can be 
taken to be proportional to comoving volume of $\Sigma_0$ 
at reheating  
\bea
V_* = 4\pi\tilde{a}^3(\tau_*)\int_0^{\xi_c}\! 
\sinh^2\xi\, d\xi \,,
\label{thermovolume}
\eea         
where $\tau_*$ is the proper time of reheating.  Note that 
the bubble scale factor at proper time $\tau_i$ is 
$\tilde{a}(\tau_i)\approx \frac{1}{2}H_i^{-1}e^{H_i\tau_i}$ 
--- this is Eq.~(\ref{sftau}) in the limit $H_i\tau\gg 1$.  
Thus Eq.~(\ref{thermovolume}) can be written
\bea
V_* = \frac{\pi}{2}\, H_i^{-3}e^{3(H_i\tau_i+N_e)}
\int_0^{\xi_c}\! \sinh^2\xi\, d\xi \,.
\label{thermovolume2}
\eea     

In Section~\ref{sec:omega} we take interest in the volume at 
thermalization, $V_*$, as well as the curvature parameter 
$\Omega_k$, evaluated on the  hypersurface $\Sigma_0$, as a 
function of $N_e$.  The curvature parameter at $\tau_0$ can 
be related to its value at any previous (proper) time using  
\bea
\Omega_k = \Omega_k^q\,(\tilde{a}_qH_q/\tilde{a}_0H_0)^2 \,,
\label{1mO1}
\eea
where the subscript 0 denotes quantities evaluated at $\tau_0$, 
and $q$ denotes quantities evaluated at some previous time.  
We set the previous time to be that of bubble nucleation, 
$\tau=0$ in open FRW coordinates.  From Eqs.~(\ref{EFEpocket}) 
and ~(\ref{sftau}), we see $\tilde{a}(\tau)H(\tau)\to 1$ and 
$\Omega_k(\tau)\to 1$ as $\tau\to 0$.  During inflation the 
scale factor expands exponentially with time, while after 
inflation it is convenient to write the scale factor as a 
function of the temperature $T$, as opposed to the proper time 
$\tau$.  Assuming instantaneous reheating and conserving 
entropy in comoving volumes, the scale factor at temperature 
$T$ can be written
\bea
\tilde{a}(T)=\frac{1}{2}H_i^{-1}e^{H_i\tau_i+N_e}
\left(\frac{g_*T_*^3}{gT^3}\right)^{\!1/3} ,
\label{sfT}
\eea
where $T_*$ is the reheating temperature and $g$ counts the 
effective number of degrees of freedom in thermal equilibrium 
($g_*$ being the corresponding quantity at the reheating 
temperature).  We neglect $H_i\tau_i$ next to $N_e$ in the 
exponent of Eq.~(\ref{sfT}), which allows us to write   
\bea
\Omega_k = \left(\frac{2H_i\,g^{1/3}_0T_0}
{H_0\,g_*^{1/3}T_*}\right)^{\!2}e^{-2N_e} \,.
\label{Omegaexpress}
\eea

To proceed, we make educated guesses at the unknown 
parameters in Eq.~(\ref{Omegaexpress}).  First, note that 
according to our assumption of instantaneous reheating, the 
Hubble rate and temperature at reheating are related by 
$H_*^2= (8\pi^3G/90)\,g_*T_*^4$.  We consider $H_i$ to be a 
factor of a few larger than $H_*$, take $g_*$ to be on the 
order of a hundred, and guess $T_*\approx 10^{-4}G^{-1/2}$ 
(i.e. GUT-scale reheating).  Putting all this together gives
\bea
\Omega_k \approx  e^{123-2N_e} \,,
\label{OmegaN}
\eea
where we have also input the present temperature 
$T_0=2.34\times 10^{-4}$ eV and Hubble rate 
$H_0=1.53\times 10^{-33}$ eV.  We comment on the effect of 
changing our guess of $T_*$ at the end of 
Section~\ref{sec:omega}.

\subsection{Structure Formation in an Open FRW Universe}
\label{sec:structure}

Anthropic selection in favor of structure formation may be 
an important effect modulating the distribution of 
$\Omega_k$.  Therefore, we take interest in the details of 
structure formation in universes in which $\Omega_k$ may 
deviate significantly from zero (the work here builds upon 
that of Refs.~\cite{VW,GTV,FKMS}).  In this section, we 
describe the relevant aspects of structure formation by 
looking at the evolution within a single bubble like ours.  
In Section~\ref{sec:omega}, we incorporate these results 
into the complete picture involving a diverging number of 
bubbles that nucleate throughout the evolution of the 
multiverse.  

In the context of estimating anthropic selection for 
structure formation, one often studies the asymptotic 
collapse fraction. This is because one is interested in 
{\it explaining}, say, the observed value of $\Lambda$, 
and one anticipates that observers like us could arise at 
times somewhat different than the present cosmic time, 
and in galaxies with mass somewhat different than that of 
the Milky Way (see for example Refs.~\cite{MSW,TARW}).  
If one were instead interested in the best possible 
{\it prediction} of $\Lambda$, one would use as much 
information as is relevant to constrain it~\cite{GV08}.  
In this case, we would take interest in the fraction of 
matter in halos with mass similar to that of the Milky 
Way, since it is in this halo that we perform the 
measurement. 

We denote the collapse fraction into halos of mass 
greater than or equal to $M_G$ at time $\tau$ as 
$F_{\rm c}(M_G,\tau)$.  The collapse fraction into only 
halos of mass equal to $M_G$ is better known as the mass 
function (evaluated at $M_G$), and we denote this 
$F_{\rm m}(M_G,\tau)$.  The collapse 
fraction $F_{\rm c}$ can be approximated using the 
Press-Schechter formalism~\cite{PS}, which gives
\bea
F_{\rm c} = {\rm erfc}\!\left[
\frac{\delta_c}{\sqrt{2}\,\sigma_{\rm rms}(M_G,\tau)}\right] \,.
\label{collapsefraction}
\eea     
Here $\delta_c$ is the collapse density threshold --- the 
amplitude reached by the linear evolution of an overdensity 
at the time when a non-linear analysis would reveal that 
it has collapsed --- and $\sigma_{\rm rms}(M_G,\tau)$ is 
the root-mean-square (rms) density contrast on the comoving 
scale enclosing a mass $M_G$ and evaluated at proper time 
$\tau$.  The collapse density threshold $\delta_c$ is not 
constant in time when $\Omega_k\ne 0$, nor when 
$\Lambda\ne 0$; however it changes by less than 10\% over 
the course of big bang evolution~\cite{GTV,PTV} and the 
collapse fraction $F_{\rm c}$ (as well as the mass function 
$F_{\rm m}$) is well-approximated by taking $\delta_c=1.69$.    

According to the Press-Schechter formalism, the mass 
function $F_{\rm m}$ can be obtained by differentiation, 
$F_{\rm m}=(dF_{\rm c}/d\ln M_G)$ --- this corresponds to 
the distribution of halo masses at any given time.  Note 
that the only $M_G$ dependence of $F_{\rm c}$ comes from 
$\sigma_{\rm rms}$.  Meanwhile, the $M_G$ dependence of 
$\sigma_{\rm rms}$ factors out of its time evolution, i.e. 
\bea
\sigma_{\rm rms}(M_G,\tau)=\bar\sigma_{\rm rms}(M_G)\,G_\Omega(\tau) \,,
\label{bsdef}
\eea
where $\bar\sigma_{\rm rms}(M_G)$ is related to the rms 
primordial density contrast on comoving scales enclosing 
mass $M_G$.  At fixed $M_G$, $d\sigma_{\rm rms}/d\ln M_G
=(d\bar\sigma_{\rm rms}/d\ln M_G)\,G_\Omega
\propto\sigma_{\rm rms}$, and so we write
\bea
F_{\rm m} \propto \frac{1}{\sigma_{\rm rms}(M_G,\tau)}
\exp\!\left[\frac{\delta_c^2}{2\,\sigma_{\rm rms}^2(M_G,\tau)}\right] \,,
\label{massfunction}
\eea
and interpret this as the mass fraction in halos with 
mass $M_G$.  Both $F_{\rm c}$ and $F_{\rm m}$ are 
functions of $\sigma_{\rm rms}$, and so we now turn to 
describing this quantity.

The factor $G_\Omega(\tau)$ in Eq.~(\ref{bsdef}) is 
called the growth factor, and an integral expression for 
it may be obtained by analyzing the first order Einstein 
field equation~\cite{Heath}.  It is convenient to first 
define a new time variable, 
\bea
x = \rho_\Lambda/\rho_m \propto \tilde{a}^3(\tau) \,,
\label{xdef}
\eea
where $\rho_\Lambda$ is the energy density in cosmological 
constant, $\rho_m$ is the matter density, and $\tilde{a}$ 
is the scale-factor (of the open FRW coordinates of the 
bubble, see Eqs.~(\ref{openFRW}) and~(\ref{EFEpocket})).  
The growth function is then 
\bea
G_\Omega(x) \propto \sqrt{1+\frac{1}{x}+\frac{k}{x^{2/3}}}
\int_0^x \frac{y^{-1/6}\,dy}{\left(1+y+k\, y^{1/3}\right)^{3/2}} \,, \,\,
\label{D}
\eea
where the curvature term $k$ is defined by matching onto 
the Einstein field equation,
\bea
H^2 = H_\Lambda^2(1+x^{-1}+k\,x^{-2/3}) \,,
\label{EFE2}
\eea
where again $H_\Lambda^2\equiv 8\pi G\rho_\Lambda/3$.  
Thus, the curvature term $k$ is related to $\Omega_k$ by
\bea
\Omega_k = \frac{k\,x^{1/3}}{1+x+k\,x^{1/3}}  \,.
\label{kOmega}
\eea

In Eq.~(\ref{EFE2}) we have ignored the presence of radiation 
in our universe, since its effect on our analysis is negligible.  
Even with this simplification, Eq.~(\ref{EFE2}) cannot be solved 
in closed form.  Instead, the evolution of $x$ with time is 
given by the integral expression
\bea
H_\Lambda\tau = \frac{1}{3} \int_0^x 
\frac{dz}{\sqrt{z^2+z+k\,z^{4/3}}} \,.
\label{tx}
\eea 
This relation defines a function $\tau(x)$ relating the 
proper time in the bubble to the new time variable $x$.  
The function $\tau(x)$ can be obtained by numerical 
interpolation and (numerically) inverted to give 
$x(\tau)$.  

\begin{figure}[t!]
\includegraphics[width=0.4\textwidth]{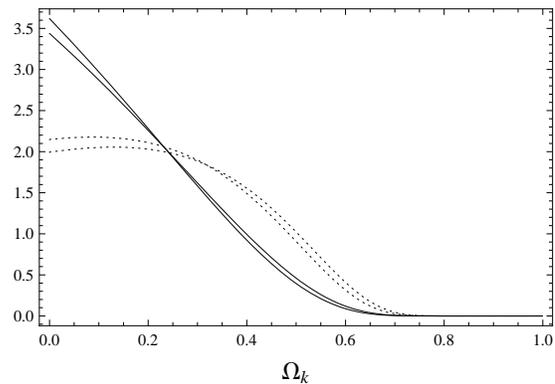} 
\caption{The collapse fraction $F_{\rm c}(\Omega_k)$ 
(solid) and mass function $F_{\rm m}(\Omega_k)$ (dotted); 
see the text for details.}
\label{fig:F}
\end{figure}

The functions $F_{\rm c}$ and $F_{\rm m}$ are in a 
sense anthropic factors, as they are approximately 
proportional to the number of observers that arise in a 
fixed comoving volume of some bubble at (or before) some 
specified time.  Note that we here use the term 
``anthropic factor'' loosely, as we are only looking at 
a single bubble and the scale-factor cutoff will introduce 
an additional selection effect when we account for all of 
the bubbles in the multiverse.  Nevertheless, it is 
worthwhile to study the distributions $F_{\rm c}(\Omega_k)$ 
and $F_{\rm m}(\Omega_k)$.  Of course, both of these depend 
on the time at which they are evaluated.  As we are 
ultimately interested in imposing a global time cutoff, we 
first evaluate $F_{\rm c}$ and $F_{\rm m}$ at a fixed 
proper time $\Delta\tau$ before the present ``time'' 
$x_0=2.88$.  The rationale behind this is to allow 
sufficient time after halo collapse for planet formation 
and the evolution of observers, while at the same time 
increasing predictive power by restricting attention to 
observers who perform the measurement of $\Omega_k$ at the 
same time that we do. 

The resulting distributions $F_{\rm c}(\Omega_k)$ and 
$F_{\rm m}(\Omega_k)$ are displayed in Fig.~\ref{fig:F}, 
where we have used $M_G=10^{12}M_\odot$, $M_\odot$ being 
the solar mass, and have chosen $\Delta\tau = 5\times 10^{9}$ 
years.  Alongside these are displayed the same 
distributions but ignoring the proper time lapse 
$\Delta\tau$, i.e. setting $\Delta\tau=0$.  We have 
normalized the distributions to integrate to unity.  Here 
and throughout the paper we use WMAP-5 mean value 
parameters~\cite{WMAP5} \footnote{The relevant values are
$\Omega_\Lambda=0.742$, $\Omega_m=0.258$, $\Omega_b=0.044$,
$n_s=0.96$, $h=0.719$, and $\Delta_{\mathcal R}^2(k=0.02\,{\rm
Mpc}^{-1})= 2.21\times 10^{-9}$.} and
compute the rms density contrast 
on the scale $M_G$ using Ref.~\cite{cmbfast} and the CMBFAST program.  
For both $F_{\rm c}$ and $F_{\rm m}$, the curve with $\Delta\tau=0$ 
is the one that is slightly higher at larger $\Omega_k$.  
Note that the distributions do not depend significantly 
on the choice of $\Delta\tau$.  For this reason, and 
because it dramatically simplifies our calculations, 
henceforth we set $\Delta\tau = 0$.

Fig.~\ref{fig:F} reveals that, although anthropic 
selection prevents an observer from measuring too large a 
value of $\Omega_k$, it does not select values of $\Omega_k$ 
as small as the observational bound ($\Omega_k\leq 0.013$ at 
95\% confidence level~\cite{WMAP5}) much more strongly than 
it selects values, say, ten times larger than this.  We 
return to this point in Section~\ref{sec:anth}.

\boldmath
\section{The Distribution of $\Omega_k$}
\label{sec:omega}
\unboldmath

We can now describe what value of $\Omega_k$ we might expect 
to measure, given certain assumptions about the multiverse.  
In any given bubble, the value of $\Omega_k$ is a function of 
the expansion history along the comoving geodesic passing 
through the spacetime point at which $\Omega_k$ is measured.  
This expansion history is well-understood only during (a 
portion of) the big bang evolution following reheating.  
Although many factors contribute to the expansion history 
before this big bang evolution, we bundle our ignorance into 
a single parameter:  the number of e-folds of slow-roll 
inflation in our bubble, $N_e$.  This is to say, we make 
guesses at relevant quantities such as the scale of inflation 
and the reheating temperature (see the end of 
Section~\ref{sec:geometry}), and consider that our errors 
are offset by (small) changes in the number of e-folds $N_e$.  
The distribution of $N_e$ is of course crucial to the 
analysis, yet in this aspect of the calculation that we must 
rely on a high degree of speculation.   

As indicated from the onset of this paper, we consider our 
universe to be a thermalized bubble in an eternally inflating 
multiverse.  Furthermore, we consider the multiverse to be 
populated by a landscape of vacua so large that we may 
consider the early dynamics of our bubble as independent of 
the low-energy physics that describes the subsequent big bang 
evolution.  In this picture, we expect the value of $N_e$ 
in our bubble to be typical of values across the multiverse, 
modulo any selection effects.  To guess at this distribution, 
we follow Freivogel, Kleban, Rodriguez Martinez, and 
Susskind (FKRMS)~\cite{FKMS}.  

These authors  consider the dominant contribution to 
$N_e$ to come 
from the slow-roll of a single scalar field over an 
approximately linear potential, 
\bea
V(\varphi) \approx V_0\left( 1-\frac{y}{\Delta}\varphi 
\right)\,, \quad \varphi_i \leq \varphi \leq \varphi_f \,,
\label{Vguess}
\eea
where $V_0$, $y$, and $\Delta\equiv\varphi_f-\varphi_i$ 
are free parameters that are assumed to scan across the 
landscape, taking values between zero and one (in Planck 
units) with uniform probability distribution.  The primordial 
density contrast can be calculated from Eq.~(\ref{Vguess}), 
and is a function of the parameters $V_0$, $y$, and $\Delta$.  
Since the primordial density contrast is known, we consider 
the slice of the landscape of $V(\varphi)$ for which it is 
fixed to the value we measure.  The resulting distribution 
of $N_e$ is~\cite{FKMS}
\bea
dP_0(N_e) \propto N_e^{-4}\, dN_e \,,
\label{Ndist}
\eea  
where here the subscript ``0'' emphasizes that we have not 
yet accounted for all of the selection effects.  
Eq.~(\ref{Ndist}) is converted into a distribution of 
$\Omega_k$ using Eq.~(\ref{OmegaN}), which gives
\bea
\frac{dP_0(\Omega_k)}{d\ln\Omega_k} \propto 
\left[61.5-\frac{1}{2}\ln\Omega_k\right]^{\!-4} 
\equiv f(\Omega_k) \,.\,\,
\label{Odist}
\eea

We now take into account the other selection effects, namely 
the effect of the scale-factor measure over the multiverse 
and the effect of anthropic selection in favor of structure 
formation.  Let us first write the result in a somewhat 
cumbersome form, in order to explain the various contributions, 
and then simplify it.  The distribution of $\Omega_k$ can be 
written
\bea
\frac{dP(\Omega_k)}{d\ln\Omega_k} &\propto& 
\lim_{t_c\to\infty}\,\int_{-\infty}^{t_c}\!
e^{\gamma t_{\rm nuc}}dt_{\rm nuc} \nn\\
&& \quad\,\times\,\int_0^{\xi_c}\!e^{3(H_i\tau_i+N_e)}
\sinh^2\xi\,d\xi \nn\\ 
&& \quad\,\times\,\,
F_{\rm m}(M_G,x_0)\,f(\Omega_k) \,. 
\phantom{\bigg(\bigg)}\quad
\label{master0}
\eea
The integral on the second line is proportional to the total 
amount of matter on the hypersurface $\Sigma_0$, given by 
Eq.~(\ref{thermovolume2}), while the mass function $F_{\rm m}$ 
selects for the fraction of matter than has collapsed into 
halos of mass $M_G$.  Collectively, these terms are 
proportional to the number of observers like us in bubbles 
that nucleate at scale-factor time $t_{\rm nuc}$ (the 
dependence on $t_{\rm nuc}$ is in the limit of integration 
$\xi_c$, see Eq.~(\ref{xic})).  The first line of 
Eq.~(\ref{master0}) integrates over all bubble nucleation 
times $t_{\rm nuc}$, with the appropriate volume weighting 
coming from eternal inflation with the scale-factor measure, 
see for example Eq.~(\ref{dn}).  This integration ignores the 
very small probability that a given vacuum might decay during 
slow-roll inflation or big bang evolution up to the 
hypersurface $\Sigma_0$.  Finally, the last term in the last 
line of Eq.~(\ref{master0}) gives the distribution of 
$\Omega_k$ coming from the dependence on the number of 
e-folds of slow-roll inflation, Eq.~(\ref{Odist}).  

As explained in Section~\ref{sec:structure}, we here use the 
mass function $F_{\rm m}$ instead of the collapse fraction 
$F_{\rm c}$ because we are interested in making a prediction, 
so we include as much relevant information as is practical --- 
in this case that we live in a halo with mass equal to that
of the Milky Way.  Thus, we set $M_G=10^{12}M_\odot$.  
Similarly, we evaluate the mass function at the present ratio 
of energy density in cosmological constant to that in matter, 
$x_0=2.88$.\footnote{We should include a time lapse 
$\Delta\tau$ to allow for planet formation and biological
evolution after halo collapse.  However, as mentioned in 
Section~\ref{sec:structure}, this complicates the analysis
but does not significantly affect the results, so for 
simplicity we neglect it.}  
One might wonder how the prediction of $\Omega_k$ is affected 
if we do not so strongly condition the calculation.  We return 
to this question in Section~\ref{sec:anth}.

To proceed, we first evaluate the inside integral over $\xi$.  
Note that all of the dependence on $\xi$ is in the factor 
$\sinh^2\xi$.  The integration can be performed analytically, 
\bea
\int_0^{\xi_c} \sinh^2\xi\,d\xi = \sinh\left(2\xi_c\right)
- 2\xi_c \,,
\eea  
with $\xi_c$ given by Eq.~(\ref{xic}).  It is convenient 
perform a variable redefinition, 
\bea
z=t_c-t_{\rm nuc}-H_0\tau_0-N_e-N_O \,,   
\eea
and exchange integration over $t_{\rm nuc}$ for integration 
over $z$.  The integration over $z$ just gives a constant 
prefactor (here and below we use $\gamma=3$).  Dropping the 
other constant factors, Eq.~(\ref{master0}) becomes
\bea
\frac{dP(\Omega_k)}{d\ln\Omega_k} \propto 
F_{\rm m}(M_G,x_0)\, f(\Omega_k) \,. 
\label{master1}
\eea
Note that Eq.~(\ref{master1}), which includes the effect of 
the scale-factor cutoff, is exactly what one would naively 
expect if the issue of measure were ignored.

\begin{figure}[t!]
\includegraphics[width=0.4\textwidth]{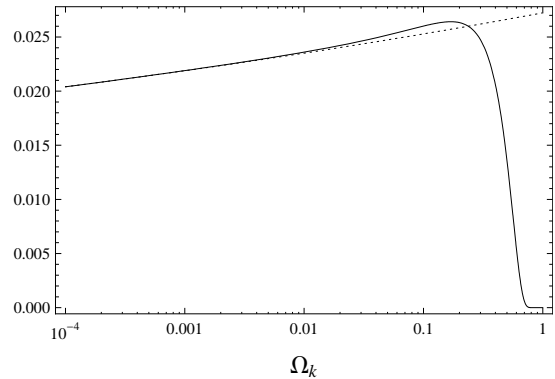} 
\caption{The relevant portion of the distribution of 
$\Omega_k$ (solid), along with a simple approximation, 
Eq.~(\ref{masterapprox}) (dotted).}
\label{fig:prediction}
\end{figure}

The distribution Eq.~(\ref{master1}) is displayed (in part) 
in Fig.~\ref{fig:prediction}.  Interestingly, the 
distribution is quite flat all the way up to rather large 
values of $\Omega_k$, falling off at $\Omega_k\approx 0.6$.  
We know from CMB measurements that 
$\Omega_k\leq 0.013$~\cite{WMAP5} (for simplicity we take 
this 95\% confidence level to be a hard bound), so to 
produce the best prediction we should cut the distribution 
at that value.  The distribution in 
Fig.~\ref{fig:prediction} is normalized as if this cut 
were in place, and the small amplitude of the distribution 
($\sim 0.02$) indicates that it has broad support over 
values of $\Omega_k$ much smaller than those displayed in 
the figure.  This can also be seen by examining the 
approximation,
\bea
\frac{dP(\Omega_k)}{d\ln\Omega_k} \sim 
\left[61.5-\frac{1}{2}\ln\Omega_k\right]^{\!-4} \,.
\label{masterapprox}
\eea
which (after proper normalization) is very accurate for 
small $\Omega_k$.  As another illustration of how broad is 
the distribution, note that the median sits at about 
$10^{-16.5}$ (corresponding to about 80 e-folds of 
slow-roll inflation).

Because of the broad support of the distribution 
Eq.~(\ref{masterapprox}), it is most likely that $\Omega_k$ 
is dominated by cosmic variance --- which is of order 
$10^{-5}$~\cite{WZ} --- instead of the relic contribution 
calculated above.  Nevertheless, it is exciting that the 
distribution of $\Omega_k$ leaves reasonable hope for 
future detection.  In particular, there is a 6\% chance to 
measure $\Omega_k\geq 10^{-3}$, and an 11\% chance to 
measure $\Omega_k\geq 10^{-4}$ (both of these percentiles 
are calculated using a distribution cut off at 
$\Omega_k=0.013$).  These results are in agreement with 
the estimates made in Ref.~\cite{FKMS}.

Recall that our analysis guessed at certain cosmological 
parameters, for example the reheating temperature, which 
was set at $T_*\approx 10^{-4}G^{-1/2}$ (c.f. the end of
Section~\ref{sec:structure}).  As a quick check of the 
effect of our guesses, consider a very different guess 
at the reheating temperature, $T_*\approx 10^{-16}G^{-1/2}$ 
(corresponding to TeV-scale reheating ).  For simplicity 
we keep our other guesses fixed.  In this case, 
the quantity ``123'' appearing in Eq.~(\ref{OmegaN}) 
becomes about 68.  Performing an analysis analogous to 
that above, we find there is a 10\% chance to measure 
$\Omega_k\geq 10^{-3}$, and an 18\% chance to measure 
$\Omega_k\geq 10^{-4}$.  Decreasing $T_*$ shifts the 
distribution of $\Omega_k$ toward larger values, but 
apparently the effect is not very strong.  The most 
important factor determining our expectations for 
$\Omega_k$ is the distribution of $N_e$ over the 
landscape.

\boldmath
\section{Anthropic Considerations and the ``Prior'' 
Distribution of $N_e$}
\label{sec:anth}
\unboldmath

The calculation of the last section was made in the spirit 
of a prediction, and as such it was conditioned by certain 
details about our location in our universe, namely that we 
inhabit a galaxy with Milky Way mass and perform our 
measurement at $x_0=2.88$.  Taking a different perspective, 
we can ask under what conditions can the landscape picture 
explain why the curvature is observed to be as small as it 
is, $\Omega_k\leq 0.013$.  In this case, we consider 
ourselves observers belonging to a more general reference 
class, and ask what values of $\Omega_k$ typical observers 
in this reference class measure.  We here consider the 
more general reference class to be observers in bubbles 
with the same low-energy physics as ours, and in galaxies 
like the Milky Way, however these observers can arise at 
any time over the course of bubble evolution.

To proceed in analogy to the calculation of 
Section~\ref{sec:omega} introduces a number of unnecessary 
complications.  Instead, we follow the methods introduced 
in Ref.~\cite{DSGSV1}.  Specifically, we take as our 
``reference objects'' not entire bubbles, but small patches 
of comoving volume, whose transverse boundaries are bubble 
walls (or the cutoff hypersurface at scale-factor time 
$t_c$).  If these patches are sufficiently small in 
spacelike extent, they may be chosen so that both 
scale-factor time $t$ and proper time $\tau$ are nearly 
constant over slicings of the patch.  These patches, like 
any reference object, arise in the multiverse at a rate 
that scales like that of Eq.~(\ref{dn}).  Integrating 
over these patches is equivalent to taking as the 
reference objects entire bubbles (cut off at $t_c$), and 
integrating over bubble nucleation times, as was done in 
Section~\ref{sec:omega}.  

The curvature parameter $\Omega_k$ is a function of the 
FRW proper time $\tau$ inside each bubble.  Therefore, to 
calculate what values of $\Omega_k$ typical observers 
measure, one must know the density of these observers as 
a function of time.  Alternatively, one can define a 
time-invariant quantity $k$, related to $\Omega_k$, and 
count the total number of observers inhabiting bubbles 
with different values of $k$.  We use
\bea
k=\left(\frac{\Omega_k^3}{\Omega_\Lambda\Omega_m^2}\right)^{1/3} \,,
\label{kdef}
\eea   
which corresponds to the quantity $k$ used in 
Eq.~(\ref{EFE2}).  Note that the observational bound 
$\Omega_k\leq 0.013$ corresponds to 
$k\leq 0.035$~\cite{WMAP5}.  

To begin the calculation, consider a spacetime volume 
that is bound from below by a small patch of some bubble 
wall at scale-factor time $t_{\rm w}$.  The number of 
observers in this volume is proportional to the collapse 
fraction evaluated at a proper time cutoff $\tau_c$, where 
$\tau_c$ is defined by the relation
\bea
t_c-t_{\rm w} = N_e + \int_{\tau_*}^{\tau_c}\! H(\tau)\,d\tau 
= N_e + \ln\!\left[\frac{\tilde{a}(\tau_c)}{\tilde{a}(\tau_*)}\right] ,
\label{tauc}
\eea  
where $\tau_*$ is (proper) time of reheating and $N_e$ is 
the number of e-folds of expansion between the bubble wall 
and reheating.  As our notation indicates, we assume the 
latter expansion comes entirely from slow-roll inflation; 
i.e. we neglect the contribution coming from the initial 
curvature-dominated phase.  

The number of observers in such a patch can then be 
approximated as proportional to
\bea
e^{3N_e}F_{\rm c}(M_G,\tau_c) \,,
\eea
where the exponential gives the volume expansion factor 
coming from slow-roll inflation, and the second term 
evaluates the collapse fraction at the proper time cutoff.  
The collapse fraction counts matter collapsed into halos 
of mass $M_G$ or greater; however halos with mass greater 
than $M_G$ at time $\tau_c$ had mass equal to $M_G$ at 
some time $\tau<\tau_c$, so these halos contribute to our 
reference class.  As we have already noted, one might 
instead evaluate $F_{\rm c}$ at some time $\Delta\tau$ 
before $\tau_c$, in order to give time for galaxies and 
observers to evolve between the time of collapse and the 
proper time cutoff.  However, including this effect 
significantly complicates the calculation, whereas in 
Section~\ref{sec:geometry} we found that it does not 
significantly affect the collapse fraction.  Therefore, 
we here neglect it.

Summing over all patches gives 
\bea
\frac{dP(k)}{d\ln k} \propto \lim_{t_c\to\infty}
\int_{-\infty}^{t_c}\! e^{3N_e+\gamma t_{\rm w}}
F_{\rm c}(M_G,\tau_c)\,\tilde{f}(k)\,
dt_{\rm w} \,, \,\,\,
\label{masterB0}
\eea
where as before we have neglected the small probability 
that a given vacuum may decay during slow-roll inflation 
or during big bang evolution.  As was the case with 
Eq.~(\ref{master0}), the exponential dependence on $N_e$ 
is an illusion.  Note that the cutoff $\tau_c$ 
corresponds to a cutoff $x_c$, where as before 
$x\equiv\rho_\Lambda/\rho_m$.  Eq.~(\ref{tauc}) gives 
$\ln x_c = 3(t_c-t_{\rm w}-N_e) + {\rm const}$, which 
can be used to change the variable of integration from 
$t_{\rm w}$ to $x_c$. This gives
\bea
\frac{dP(k)}{d\ln k} \propto
\int_0^{\infty}\! x_c^{-2} F_{\rm c}(M_G,x_c)\,
\tilde{f}(k)\, dx_c \,,
\label{masterB1}
\eea 
where we have used $\gamma=3$.  Note that the ``prior'' 
distribution $\tilde{f}(k)$ factors out of the 
integration.   

The argument of Eq.~(\ref{masterB1}) contains a factor 
of $x_c^{-2}$.  This factor induces the ``youngness bias'' 
of the scale-factor cutoff measure, which prefers bubbles 
that nucleate nearer to the cutoff (for which $x_c$ is 
smaller).  As shown in Ref.~\cite{DSGSV1}, this bias is 
rather mild.  It does not appear in the calculation of 
Section~\ref{sec:omega}, c.f. Eq.~(\ref{master1}), 
because that calculation was performed at fixed $x$, 
$x=x_0$.    

Whereas $f(\Omega_k)$ of Eq.~(\ref{master1}) corresponds 
to the distribution of $\Omega_k(x)$ at fixed $x=x_0$, 
the function $\tilde{f}(k)$ of Eq.~(\ref{masterB1}) 
corresponds to the distribution of $k$, which is 
independent of $x$.  Using 
$T\propto 1/\tilde{a} \propto x^{-1/3}$ and 
Eqs.~(\ref{Omegaexpress}) and~(\ref{kdef}), we find
\bea
k = e^{124-2N_e}\,,
\eea
where the additional factors of $\Omega_\Lambda$ and 
$\Omega_m$ essentially change the ``123'' of 
Eq.~(\ref{OmegaN}) to ``124.''  In the case 
$dP_0(N_e)\propto N_e^{-4}dN_e$, this gives the 
distribution 
\bea
\tilde{f}(k) \equiv \frac{dP_0(k)}{d\ln k}
\propto \left[62-\frac{1}{2}\ln k\right]^{-4} . 
\label{tfdef}
\eea      

\begin{figure}[t!]
\includegraphics[width=0.4\textwidth]{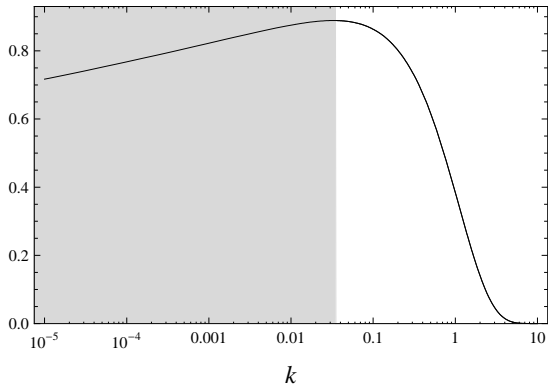} 
\caption{The distribution $dP(k)/dk$; see the text for 
details.  The present observationally acceptable region, 
$k\leq 0.035$, is indicated by shading.}
\label{fig:anth}
\end{figure}

In Fig.~\ref{fig:anth} we display $dP(k)/dk$, using 
Eq.~(\ref{masterB1}) with $\tilde{f}(k)$ given by 
Eq.~(\ref{tfdef}).  As in Fig.~\ref{fig:prediction}, we 
have cropped the figure to more clearly illustrate the 
region of interest (the cropped portion very closely
follows the distribution $dP_0(k)/dk$).  The 
observationally acceptable region, $k\leq 0.035$, is 
indicated by shading.  Clearly, values of $k$ satisfying 
our observational bound are not atypical in the FKRMS 
landscape model of inflation; in fact 93\% of observers 
measure $k$ to satisfy this bound.       

\begin{figure}[t!]
\includegraphics[width=0.4\textwidth]{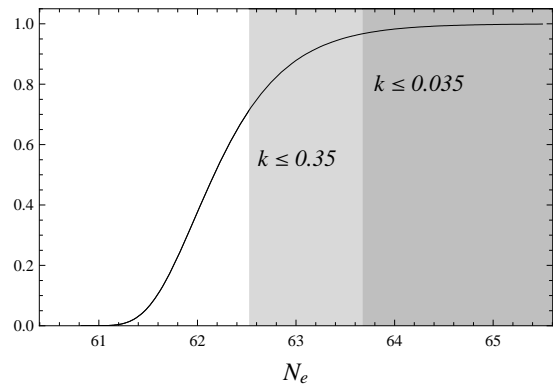} 
\caption{The distribution $dP(N_e)/dN_e$ assuming a 
flat ``prior'' for $N_e$, i.e. $dP_0(N_e)/dN_e=$ 
constant.  The shaded regions correspond to the 
observatinal bound $k\leq 0.035$, and a bound ten 
times larger, translated to number of e-folds.}
\label{fig:anth2}
\end{figure}

Although typical observers measure $k\leq 0.035$, note
that anthropic selection for structure formation, which 
causes the distribution of $k$ to fall off at large $k$,
does not select for values of $k$ satisfying the 
observational bound much more strongly than it selects 
for values, say, ten times larger.  This is more clearly 
illustrated if we plot the distribution $dP(N_e)/dN_e$ 
--- i.e. the distribution of the observed number of 
e-folds $N_e$ --- using a flat ``prior'' for $N_e$, in 
other words setting $dP_0(N_e)/dN_e=$ constant.  This is 
done in Fig.~\ref{fig:anth2}.  The observational bound 
on $k$, and a bound ten times larger, are converted to 
e-folds of inflation and represented by the shaded 
regions.    

A flat prior for $N_e$ is unrealistic, but it serves to 
illustrate the effect of anthropic selection.  As 
expected, the distribution of $N_e$ is exponentially 
suppressed for small values of $N_e$, where 
Fig.~\ref{fig:anth2} reveals that in this context 
``small'' means $N_e\lesssim 61$.  The present 
observational bound, $k\leq 0.035$, corresponds to 
$N_e\geq 63.7$.  Although the lower limit of this bound 
is not much larger than the anthropic cutoff at 
$N_e\approx 61$, $k$ depends exponentially on $N_e$, and 
we can see from Fig.~\ref{fig:anth2} that values of $k$ 
over ten times larger than the present bound are not 
strongly suppressed.  This, in principle, allows us to 
exclude hypothetical landscape models of inflation, 
based on their predicting $k$ to be larger than the 
observational bound.

In particular, if a hypothetical model of inflation in the 
landscape predictions a distribution of $N_e$ that too 
strongly prefers smaller values of $N_e$, then it is 
possible for us to exclude this model based on the 
measurement $k\leq 0.035$.  This is enticing because models 
of inflation in string theory tend to prefer a smaller 
number of e-folds of slow-roll inflation.  On the other 
hand, it is important to recognize that in order to exclude 
a landscape model of inflation, we require a certain 
``fortuitous'' shape to the ``prior'' distribution 
$dP_0(N_e)/dN_e$.  For example, many classes of potentials 
will strongly prefer smaller values of $N_e$ when $N_e$ is 
small, but this region of the parameter space is not 
relevant to our observations, because values $N_e\lesssim 61$ 
are exponentially suppressed and do not contribute to the 
full distribution $dP(N_e)/dN_e$.  
   
Let us illustrate this with an example.  Consider a 
landscape model of inflation that predicts a power-law 
prior distribution of $N_e$, 
\bea
dP_0(N_e) \propto N_e^{-\alpha}dN_e \,.
\label{powerlaw}
\eea
According to our assumptions, such a distribution is ruled 
out at greater than 95\% confidence level when 
$\int_0^{0.035}(dP(k)/dk)\,dk<0.025$, where $dP(k)/dk$ is 
here presumed to be normalized to unity.  Performing the 
calculation, we find $\alpha\geq 114$ is ruled out.  That only 
such a strong power-law dependence can be ruled out may be 
striking, but it is easy to understand in light of our 
above remarks.  The observationally excluded region is, 
roughly speaking, $N_e<64$; however anthropic selection 
suppresses all contributions from the interval $N_e<61$.  
Therefore a landscape model of inflation is ruled out only 
if the prior distribution $dP_0(N_e)/dN_e$ has much more 
weight in the interval $61\leq N_e \leq 64$ than in the 
interval $64<N_e$.  In the context of the power-law 
distribution of Eq.(\ref{powerlaw}), we require
\bea
1\ll \frac{\int_{61}^{64}N_e^{-\alpha}dN_e}
{\int_{64}^{\infty}N_e^{-\alpha}dN_e} \approx
\frac{3(\alpha-1)}{64} \,.
\label{samplecalc}
\eea  
Thus, roughly speaking, we expect to rule out power-law 
distributions only if $\alpha \gg 20$.  The large power is
explained by the fact that the prior distribution must have
sharp behavior at large values of $N_e$.  

Although it is hard to imagine how a landscape model of 
inflation could give such a strong power-law prior 
distribution of $N_e$, it is not implausible that a more 
realistic model of inflation, which could give a much more 
complicated prior distribution of $N_e$, could have the 
necessary sharp behavior at large $N_e$.  Let us note, for 
instance, that potential energy barriers --- as are 
necessary in the bubble-nucleation model we are considering 
--- will give a sharp cutoff at large $N_e$.  
  
Finally, we emphasize that the above analysis, which refers
specifically to numbers like $N_e=63.7$, etc., relies 
implicitly on a number of assumptions in addition to the 
form of the prior distribution of $N_e$, for example the 
reheating temperature.  These are described at the end of
Section~\ref{sec:geometry}.  Yet, different assumptions 
would merely shift the specific values of $N_e$ mentioned 
above, and our conclusions would be unchanged.

\section{Conclusions}
\label{sec:conclusions}

Our universe may be contained in one among a multitude of 
diverse bubbles in an eternally inflating multiverse.  If 
the fundamental theory permits a landscape of meta-stable 
solutions, then in general one cannot make hard 
predictions about aspects of our universe, but must 
instead understand it via conditional probability 
distributions.  In such a multiverse, a diverging number 
of infinite-volume bubbles are formed, and weighing the 
various possibilities against each other requires 
regulation of these divergences.  Different regulators 
give different observational predictions, and in this 
work we study the distribution of the curvature parameter 
$\Omega_k$ using one of the most promising regulators, 
the scale-factor cutoff.   

In a large landscape, the vacuum of our bubble might be 
reached by tunneling from a number of different ``parent'' 
vacua.  Then, depending on in which parent vacuum our 
bubble nucleates, we in general expect different early 
universe dynamics, including different possibilities for
the number of e-folds of slow-roll inflation $N_e$.  In a 
very large landscape, as is expected from string theory, 
we also expect a large number of vacua with low-energy 
physics indistinguishable from our own.  In this case, 
one expects a smooth distribution of possible values of 
$N_e$ describing our bubble.  One of the features of the 
scale-factor cutoff measure is that it does not reward 
bubbles for having a longer duration of slow-roll 
inflation.  This raises the possibility that $N_e$ may 
not be too much larger than is needed to pave the way for 
structure formation, and therefore that $\Omega_k$ may 
be large enough to distinguish from the value expected 
from cosmic variance, $\sim 10^{-5}$.

Freivogel, Kleban, Rodriguez Martinez, and Susskind 
(FKRMS) have proposed a toy model of inflation in the 
landscape, which gives a ``prior'' distribution of $N_e$ 
of the form $dP_0(N_e)\propto N_e^{-4}\,dN_e$ (on the slice 
of the parameter space corresponding to a fixed primordial 
density contrast $Q$).  Using the scale-factor 
cutoff measure, we find this distribution predicts a 6\% 
chance to observe $\Omega_k\geq 10^{-3}$, and an 11\% 
chance to observe $\Omega_k\geq 10^{-4}$, thus confirming
the results of FKRMS~\cite{FKMS}.  

Although in the FKRMS model of inflation in the landscape 
observers typically measure   
$k=(\Omega_k^3/\Omega_\Lambda\Omega_m^2)^{1/3}$ to satisfy 
our observational bound, $k\leq 0.035$, anthropic 
selection does not strongly suppress values of $k$ over 
ten times larger than this (when asking what typical 
observers measure, it is convenient to refer to the 
time-independent curvature term $k$ rather than the 
time-dependent curvature parameter $\Omega_k$).  Thus, we 
may use the observed bound on $k$ to rule out hypothetical 
landscape models of inflation that too strongly prefer 
smaller values of $N_e$.

Anthropic selection is not strong in the vicinity of the 
observational bound $k\leq 0.035$, however sufficiently 
large values of $k$ are strongly suppressed.  Put another 
way, with some assumptions about inflation $k\leq 0.035$ 
corresponds to $N_e\geq 63.7$.  Anthropic selection is not 
strong in the vicinity of $N_e = 63.7$, but 
exponentially suppresses $N_e\lesssim 61$.  This is to say 
a hypothetical model of inflation that very strongly 
prefers smaller values of $N_e$ for $N_e\lesssim 61$ does 
not conflict with our observational bound, since this range 
of $N_e$ is strongly anthropically suppressed.  On the 
other hand, if a hypothetical model of inflation gives a 
prior distribution of $N_e$ that strongly prefers $N_e$ in
the interval $61\lesssim N_e < 63.7$, relative to it 
being in the interval $63.7 < N_e$, then such a model 
can be ruled out using our observational bound.

\begin{acknowledgments}
We are grateful to Alan Guth and Alex Vilenkin for 
collaboration on parts of this work, and also thank Ken Olum 
for illuminating discussions.  The work of ADS was supported 
in part by the INFN ``Bruno Rossi'' Fellowship, and in part 
by the U.S.  Department of Energy (DoE) under contract No. 
DE-FG02-05ER41360.  MPS was supported in part by the U.S. 
National Science Foundation under grant NSF 322.
\end{acknowledgments}

\appendix

\section{``Local'' scale-factor cutoff measure}
\label{sec:altSFC}

We here repeat the analysis of Sections~\ref{sec:omega} 
and~\ref{sec:anth}, but performing a cutoff on the ``local''
scale-factor time $t'$ (see Section~\ref{sec:sfcutoff}), 
where we use the prime to help distinguish the results here 
from those of the ``FRW'' scale-factor time $t$ displayed 
throughout the main text.  It is convenient to approach the 
problem in the manner of Section~\ref{sec:anth}; that is we 
take as our reference objects small patches of comoving 
volume, with transverse boundaries corresponding to bubble 
walls (or the scale-factor cutoff hypersurface at $t'_c$).  
Again, if these patches are 
sufficiently small in their spacelike extent, the 
scale-factor time $t'$ and the proper time $\tau$ are 
nearly constant over spacelike slicings of the patches.  
Analogous to in Section~\ref{sec:anth}, if we label each 
patch by the scale-factor time of reheating in the patch, 
$t'_*$, then such patches arise in the multiverse at a 
rate proportional to $e^{\gamma t'_*}$.

The scale-factor time $t'$ probes expansion on 
infinitesimal scales.  However, we take the number of 
observers to be proportional to the number of Milky 
Way--like galaxies, and we model such galaxies using
spherical top-hat overdensities with mass 
$M_G=10^{12}M_\odot$, so there is no need to probe 
scales smaller than the comoving volume that encloses 
mass $M_G$.\footnote{Realistically, structure formation 
is hierarchical:  small scales turn around and collapse 
before larger scales.  When the region surrounding a 
given geodesic collapses, its scale-factor time becomes 
frozen.  Thus, it would seem we cannot ignore structure 
formation on such small scales.  However, whether or not 
any observers arise in some small collapsed structure 
depends on whether that structure combines with others to 
form a larger structure --- ultimately a large galaxy.  
We model the requirement that small structures coalesce 
into larger ones as equivalent to requiring that structure 
formation occurs on the largest necessary scale, using a 
spherical top-hot model for the initial overdensity.}  
The probability that a comoving patch enclosing mass $M_G$ 
contains an observer is then proportional to the 
probability that such a patch begins to collapse before 
the scale-factor time cutoff $t'_c$.  (Recall that we have 
defined the scale factor cutoff such that geodesics in 
collapsing regions are extended unless or until they hit a 
singularity.)  This probability can be parametrized in 
terms of the spacetime curvature of the patch at, say, the 
reheating time $t'_*$.  

By Birkhoff's theorem, the evolution of a comoving patch 
enclosing a spherical top-hat overdensity is equivalent to 
that of a closed FRW universe with field equation
\bea
(\dot{y}/3y)^2=H_\Lambda^2(1+y^{-1}-\kappa\, y^{-2/3}) \,.
\label{EFE3}
\eea
The ``local scale factor cube root'' $y$ is defined so as 
to coincide with the ``bubble scale factor cube root'' $x$ 
of Eq.~(\ref{EFE2}) (c.f. Eq.~(\ref{xdef})) at early times.  
The total spacetime curvature $\kappa$ is the sum of the 
bubble curvature $k$ (coming from the global bubble 
geometry) and the primordial curvature perturbation $\mR$ 
(coming from quantum fluctuations during inflation).  We 
define $\mR$ to be positive for overdensities, so that
\bea
\kappa = \mR - k \,.
\eea

The spherical overdensity will turn around and begin to 
collapse before the scale-factor time cutoff $t'_c$ only 
if the curvature exceeds some minimum value 
$\kappa_{\rm min}(t'_c,t'_*)$.  For a bubble with given
value of $k$, the probability for this to occur is 
\bea
{\mathcal A}(k;t'_c,t'_*) &\propto& \int_{\kappa_{\rm min}}^\infty 
\exp\left[-\frac{(\kappa+k)^2}{\mR_{\rm rms}^2}\right]
d\kappa \nn\\
&\propto& {\rm erfc}\!\left[ \frac{\kappa_{\rm min}(t'_c,t'_*)+k}
{\sqrt{2}\,\mR_{\rm rms}}\right] \,,
\label{aanth}
\eea  
where we assume $\mR$ has a Gaussian distribution with rms 
value $\mR_{\rm rms}$.  As our notation suggests, 
${\mathcal A}$ can be interpreted as an anthropic factor, 
giving the probability that a given patch contains an 
observer.  The probability to observe a given value of $k$ 
is thus
\bea
\frac{dP(k)}{d\ln k} \propto \lim_{t'_c\to \infty}\int_{-\infty}^{t'_c}
{\mathcal A}(k;t'_c,t'_*)\,\tilde{f}(k)\,e^{\gamma t'_*}\,dt'_* \,, 
\label{aprob}
\eea
where, as in Eq.~(\ref{tfdef}), $\tilde{f}(k)$ is the 
(logarithmic) distribution of $k$ among universes with big 
bang evolution like ours, and $e^{\gamma t'_*}$ is 
proportional to the number of patches at scale-factor time 
$t'_*$.   

It is left to solve for $\kappa_{\rm min}(t'_c,t'_*)$.  
First note that a spherical overdensity described by 
Eq.~(\ref{EFE3}) turns around and begins to collapse when 
$\dot{y}=0$, or when $1+y^{-1}-\kappa\, y^{-2/3}=0$.  Thus 
we can write
\bea
\kappa(y_{\rm turn}) = y_{\rm turn}^{-1/3}
\left(1+ y_{\rm turn}\right) \,.
\eea
Meanwhile, $\kappa_{\rm min}$ is simply the value of 
$\kappa$ for which $t'_c-t'_*=(1/3)\ln(y_{\rm turn}/y_*)$, 
where $y_*$ is the local scale factor at the time of 
reheating.  (Here we use the definition of scale-factor 
time, $t'=\ln a$, along with $y\propto a^{1/3}$.)  Thus we 
can write
\bea
\kappa_{\rm min}(t'_c,t'_*) = y_*^{-1/3}e^{t'_*-t'_c}
\left[1+y_*e^{3(t'_c-t'_*)}\right] \,. 
\label{kmin}
\eea  

\begin{figure}[t!]
\includegraphics[width=0.4\textwidth]{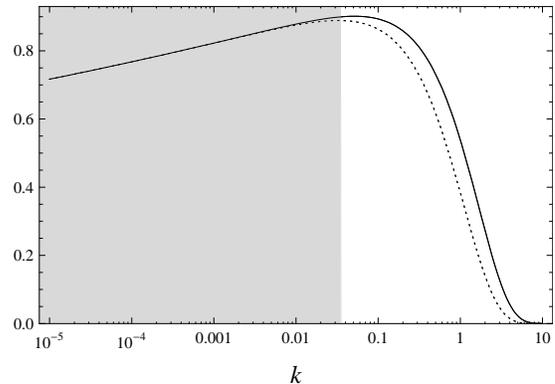} 
\caption{The distribution $dP(k)/dk$ using scale-factor 
time $t'$ (solid) and $t$ (dashed); see text for details.
The normalizations are chosen for clear comparison, while
the shaded region indicates the observed bound 
$k\leq 0.035$.}
\label{fig:app}
\end{figure}

The distribution of observed values of the bubble 
curvature $k$ is obtained by combining Eq.~(\ref{aprob}) 
with Eq.~(\ref{aanth}) and Eq.~(\ref{kmin}).  The 
resulting expression is simplified if we change the 
integration variable from $t'_*$ to  
$y_{\rm turn}=y_*^{1/3}e^{t'_c-t'_*}$.  Then we can write
\bea
\frac{dP(k)}{d\ln k} \propto \int_{y_*}^{\infty} \!\!
{\rm erfc}\!\left[ \frac{1+ky_{\rm turn}\!+y_{\rm turn}^3}
{\sqrt{2}\,\mR_{\rm rms}\,y_{\rm turn}}\right]\!
\frac{\tilde{f}(k)}{y_{\rm turn}^4}\,dy_{\rm turn}\,, \,\,\,\,
\label{masterC2}
\eea  
where we have used $\gamma=3$.  It makes no difference if 
we simply set $y_*\to 0$ in the lower limit of 
integration.  This expression corresponds to the analogue 
of Eq.~(\ref{masterB1}), but for the local scale-factor time 
$t'$, as opposed to the FRW scale-factor time $t$.  
$\mR_{\rm rms}$ is the rms primordial curvature perturbation 
on comoving scales enclosing mass $M_G$ --- it is related to, 
say, the rms density contrast $\sigma_{\rm rms}$ by
\bea
\mR_{\rm rms} = (5/3)\sigma_{\rm rms}(M_G,\tau_F)\,
x_F^{-1/3} \,,
\eea  
where the quantities on the right-hand side are evaluated 
at some fiducial time $\tau_F$ during matter domination, 
i.e. before vacuum energy or curvature become significant.  
(This relation obtains from matching the linearized 
Einstein field equation onto Eq.~(\ref{EFE3}).)  

Fig.~\ref{fig:app} displays $dP(k)/d\ln k$ using the 
scale-factor cutoff measure for both scale-factor time 
$t'$ and scale-factor time $t$.  We use Eq.~(\ref{tfdef}) 
to determine $\tilde{f}(k)$ for clear comparison, and the
shaded region indicates the bound $k\leq 0.035$.  As 
advertised in the introduction, the two definitions of 
scale-factor time give qualitatively similar results, 
however the anthropic suppression of large values of $k$
kicks in at larger $k$ when using the locally-defined 
scale-factor time $t'$.  The two distributions are very
similar for $k$ less than the observed bound, indicating
that the predictions of Section~\ref{sec:omega} are 
essentially unchanged when using the local scale-factor
time.  On the other hand, since the local scale-factor
time measure permits larger values of $k$ before strong
anthropic suppression, if this is the correct measure 
then it would be somewhat easier (than indicated in 
Section~\ref{sec:anth}) to rule out landscape models of
inflation that prefer smaller values of $N_e$.

\end{document}